\documentclass[aps,pra,twocolumn,superscriptaddress]{revtex4}
\usepackage{amssymb}
\usepackage{graphicx}
\usepackage{amsmath}

\newcommand{\vect}{\mathbf}

\begin{document}

\title{Landau Damping in a Mixture of Bose and Fermi Superfluids}
\author{Huitao Shen}
\affiliation{Institute for Advanced Study, Tsinghua University, Beijing, 100084, China}
\author{Wei Zheng}
\email{zhengwei8796@gmail.com}
\affiliation{Institute for Advanced Study, Tsinghua University, Beijing, 100084, China}
\date{\today }

\begin{abstract}
We study the Landau damping in Bose-Fermi superfluid mixture at finite
temperature. We find that at low temperature, the Landau damping rate will
be exponentially suppressed at both the BCS side and the BEC side of Fermi
superfluid. The momentum dependence of the damping rate is obtained, and it
is quite different from the BCS side to the BEC side. The relations between
our result and collective mode experiment in the recently realized
Bose-Fermi superfluid mixture are also discussed.
\end{abstract}

\maketitle

\section{Introduction}

The quasiparticle is an important concept in modern many-body physics. The
statistics and dispersion of quasiparticles determine various low energy
properties of the quantum many-body system. In the general situation, there
exsits interaction between the quasiparticles. That leads two major effects.
One is the modification of the dispersion of excitations, corresponding to
the real part of the self-energy of quasiparticle. Second, the interaction
can also damp a given excitation, i.e. giving quasiparticle a finite
lifetime, which can be reflected by the imaginary part of the self-energy.
The damping of low energy excitations is responsible for many interesting
phenomena of many-body system, such as transport and thermalization.

For example, in a uniform condensed Bose gas with short-range interaction,
the low energy excitation is phonon-like quasiparticle with linear
dispersion when wavelength is larger than the healing length. While when
wavelength is smaller than healing length, the excitation is
free-particle-like quasiparticle with quadratic dispersion. At zero
temperature, due to the residual interaction between excitations, a
quasiparticle in the BEC can be damped by decaying into two quasiparticles.
Since each product of decay must have an energy lower than the original
quasiparticle, the final state phase space is restricted. This gives the
damping rate a very sensitive dependence on the initial momentum $\gamma
\sim k^{5}$ \cite{Stringaribook}. This decay process is the so-called
Beliaev damping. At finite temperature, a given quasiparticle can also be
damped by absorbing thermal quasiparticles. As a result, it is very
sensitive to temperature, $\gamma \sim T^{4}k$ \cite{Pethick}. This
mechanism is known as Landau damping, which was first discussed in plasma
oscillation by Landau \cite{Landau}. It plays an important role in various
phenomena such as anomalous skin effect in metals and the damping of phonons
in solids. Landau-Beliaev damping in a uniform Bose superfluid has been
widely studied both theoretically \cite{Beliaev}\cite{LandauBaliaev}\cite%
{Vincent}\cite{Stringari2} and experimentally \cite{exp2}. In trapped
system, since the low energy excitations are discrete, Beliaev damping is
forbidden. The damping of low energy modes is attributed to Landau process,
and the experiment damping rate has been found to be consistent with the
theory of Landau damping \cite{exp1}\cite{Gora}. Landau-Beliaev damping has
also been studied in dipolar BECs \cite{dampdipole} and in the mixture of
BEC with normal Fermi gas \cite{dampmix}.

Recently a Bose-Fermi superfluid mixture has been first realized by ENS
group \cite{ENS}. The dipole mode of this new superfluid mixture has been
measured. It exhibits a frequency shift and an unusual damping behavior.
This experimental development triggers many investigations on Bose-Fermi
superfluid mixture \cite{pre}\cite{Stringari}\cite{ad}\cite{ZhengWei}\cite%
{rotMix}\cite{Bruun}.

The quasiparticles in this Bose-Fermi superfluid mixture have a quite unique
feature. There are two gapless bosonic modes, which are Goldstone modes of
Bose and Fermi superfluids. It also has a gapped fermionic mode,
corresponding to the Cooper pair breaking. Moreover, in the ENS experiment,
the Fermi superfluid can be tuned from the BCS side to the BEC side by
Feshbach resonance. During the crossover, the behavior of three kinds of
excitations gradually changes from the BCS limit to the BEC limit. In the
BCS limit, the velocity of Goldstone mode in Fermi superfluid is quite
large, approaching $v_{F}/\sqrt{3}$ \cite{FGoldstone}, while the gap of the
fermionic mode is exponentially small. When it is tuned to the BEC side, the
velocity of the Goldstone mode decreases monotonously \cite{RMP}\cite{Thomas}%
, while the gap becomes larger and larger. The dispersions of the three
excitations at both sides are plotted in Fig.\ref{dis}.

To understand the unusual damping behavior of dipole mode in the ENS
experiment, Zheng and Zhai study the Beliaev damping of bosonic mode in Bose
superfluid by considering its interacting with quasiparticles in Fermi
superfluid at zero temperature \cite{ZhengWei}. They found that Beliaev
process will be activated only if the excitation momentum exceeding a
critical value $k_{c}$. This threshold damping behavior, i.e. $\gamma \sim
\left( k-k_{c}\right) ^{\alpha }$, is quite different at the BCS side and
the BEC side. To be specific, at the BCS side $\alpha =0$, while at the BEC
side $\alpha =3$. This is because at the BCS side, the damping is dominated
by decaying into fermionic pair-breaking modes in Fermi superfluid. The
final state phase space of fermionic mode is restricted by density-of-state
near the Fermi surface, so that the damping rate is nearly a constant. On
the other hand, such a restriction does not exist at the BEC side, where the
damping is dominated by decaying into Goldstone modes in Fermi superfluid.
So the damping rate grows rapidly with momentum. However, they only consider
the zero-temperature case, in which Landau damping is frozen. It is nature
to ask the question, what will happen at finite temperature when Landau
damping is activated.

In this paper, we investigate Landau damping of bosonic quasiparticles in
Bose superfluid due to its interacting with quasiparticles in Fermi
superfluid at finite temperature. We consider the typical cold atom
situation, in which Fermi superfluid is in the strongly interacting regime,
while Bose superfluid is in the weakly interacting regime. We find that
unlike Beliaev damping, the critical momentum for Landau damping is zero,
i.e. $k_{c}=0$, both at the BCS side and the BEC side. We obtained the
temperature dependence of Landau damping, showing that the damping rate is
exponentially suppressed at temperature low enough at both sides. This is
indeed a departure from the $T^{4}$ dependence in the single component BEC.
The momentum dependence of the damping rate is also obtained, and its
behavior is quite different from the BCS side to the BEC side. At BCS side,
the damping rate grows linearly with momentum, while at the BEC side, it is
nearly a constant. This damping behavior revealing different dominated low
energy quasiparticles at each side.

This paper is organized as follows: In Sec. II we construct the model for
Bose-Fermi superfluid mixture and its mean field treatment. In Sec. III
Landau damping is considered by perturbation method at the BCS side. In Sec.
IV Landau damping is calculated at the BEC side. In Sec. V, our main results
are summarized and the connection to experiment is discussed.

\section{The Model}

Consider a homogeneous mixture of Bose and Fermi superfluid. The Hamiltonian
of the superfluid mixture has three parts: $\hat{H}=\hat{H}_{b}+\hat{H}_{f}+%
\hat{H}_{bf}$,
\begin{eqnarray}
\hat{H}_{b} &=&\int d^{3}\mathbf{r}\left\{ \hat{b}^{\dagger }(\mathbf{r})%
\hat{H}_{0,b}\hat{b}(\mathbf{r})+\frac{g_{b}}{2}\hat{b}^{\dagger }(\mathbf{r}%
)\hat{b}^{\dagger }(\mathbf{r})\hat{b}(\mathbf{r})\hat{b}(\mathbf{r}%
)\right\} ,  \notag \\
\hat{H}_{f} &=&\int d^{3}\mathbf{r}\left\{ \sum_{\sigma }\hat{c}_{\sigma
}^{\dagger }(\mathbf{r})\hat{H}_{0,f}\hat{c}_{\sigma }(\mathbf{r})\right.
\notag \\
&&\left. +g_{f}\hat{c}_{\uparrow }^{\dagger }(\mathbf{r})\hat{c}_{\downarrow
}^{\dagger }(\mathbf{r})\hat{c}_{\downarrow }(\mathbf{r})\hat{c}_{\uparrow }(%
\mathbf{r})\right\} ,  \notag \\
\hat{H}_{bf} &=&g_{bf}\sum_{\sigma }\int d^{3}\mathbf{r}\hat{b}^{\dagger }(%
\mathbf{r})\hat{b}(\mathbf{r})\hat{c}_{\sigma }^{\dagger }(\mathbf{r})\hat{c}%
_{\sigma }(\mathbf{r}),
\end{eqnarray}%
where $\hat{H}_{0,i}=-\hbar ^{2}\nabla ^{2}/(2m_{i})-\mu _{i}$, $i=b,f$
denotes bosons and fermions. $\sigma =\uparrow ,\downarrow $ denotes the
spin components of fermions. The coupling constants are related to the
scattering lengths: $1/g_{b}=m_{b}/(4\pi \hbar ^{2}a_{b})$ and $%
1/g_{f}=m_{f}/(4\pi \hbar ^{2}a_{f})+\sum_{\mathbf{k}}1/(2\varepsilon _{%
\mathbf{k}}^{f})$, where the scattering lengths can be tuned by Feshbach
resonance. When a magnetic field is near a Feshbach resonance between
fermions, $g_{b}$ and $g_{bf}$ are generally in the weakly interacting
regime and are approximately constant according to the experiment setup.
Therefore for $\hat{H}_{b}$, we take the Bogoliubov mean-field theory. For $%
\hat{H}_{f}$, we take the BCS-BEC crossover mean-field theory. After
mean-field treatment, one obtains
\begin{align}
\hat{H}_{b}^{\mathrm{MF}}& =\sum_{\mathbf{k}}E_{\mathbf{k}}^{b}\hat{\alpha}
_{\mathbf{k}}^{\dagger }\hat{\alpha} _{\mathbf{k}}, \\
\hat{H}_{f}^{\mathrm{MF}}& =\sum_{\mathbf{k}}E_{\mathbf{k}}^{f}(\hat{\beta}
_{\mathbf{k}}^{\dagger }\hat{\beta} _{\mathbf{k}}+\hat{\gamma} _{\mathbf{k}%
}^{\dagger }\hat{\gamma} _{\mathbf{k}}),
\end{align}%
where $E_{\mathbf{k}}^{b}=\sqrt{\varepsilon _{\mathbf{k}}^{b}\left(
\varepsilon _{\mathbf{k}}^{b}+2g_{b}n_{b}\right) }$, $\hat{\alpha} _{k}$ is
the quasiparticle operator for bosonic Goldstone mode in Bose superfluid. $%
E_{\mathbf{k}}^{f}=\sqrt{(\varepsilon _{\mathbf{k}}^{f}-\mu _{f})^{2}+\Delta
^{2}}$. $\hat{\beta} _{\mathbf{k}}$ and $\hat{\gamma} _{\mathbf{k}}$ are the
quasiparticle operators for fermionic pair-breaking mode in Fermi
superfluid. $\varepsilon _{\mathbf{k}}^{i}=\hbar ^{2}k^{2}/(2m_{i})$, $i=b,f$
denotes the kinetic energy of bosons and fermions. The corresponding
Bogoliubov transformations of these operators are given by
\begin{eqnarray}
\hat{b}_{\mathbf{k}} &=&u_{\mathbf{k}}^{b}\hat{\alpha} _{\mathbf{k}}-v_{%
\mathbf{k}}^{b}\hat{\alpha} _{-\mathbf{k}}^{\dagger },\  \\
\hat{c}_{\mathbf{k},\uparrow } &=&u_{\mathbf{k}}^{f}\hat{\beta} _{\mathbf{k}%
}+v_{\mathbf{k}}^{f}\hat{\gamma} _{-\mathbf{k}}^{\dagger }, \\
\hat{c}_{\mathbf{k},\downarrow } &=&u_{\mathbf{k}}^{f}\hat{\gamma} _{\mathbf{%
k}}-v_{\mathbf{k}}^{f}\hat{\beta} _{-\mathbf{k}}^{\dagger }.
\end{eqnarray}%
Here momentum-dependent coefficients $u_{\mathbf{k}}$ and $v_{\mathbf{k}}$
are given by
\begin{align}
u_{\mathbf{k}}^{b}(v_{\mathbf{k}}^{b})& =\sqrt{\frac{1}{2}\left( \frac{%
\varepsilon _{\mathbf{k}}^{b}+g_{b}n_{b}}{E_{\mathbf{k}}^{b}}\pm 1\right) },
\\
u_{\mathbf{k}}^{f}(v_{\mathbf{k}}^{f})& =\sqrt{\frac{1}{2}\left( 1\pm \frac{%
\varepsilon _{\mathbf{k}}^{f}-\mu _{f}}{E_{\mathbf{k}}^{f}}\right) }.
\end{align}

For fermionic pair-breaking mode in Fermi superfluid, when $-1/k_{F}a_{f}$
gets smaller from the BCS side to the BEC side, $\Delta $ will increase and $%
\mu _{f}$ will decrease. Apart from the pair-breaking mode, there is also a
bosonic mode of the center-of-mass motion of Cooper pairs in Fermi
superfluid, which is beyond BCS-BEC mean-field theory. Its dispersion
relation at low energy is linear: $E_{\mathbf{k}}^{m}=\hbar c_{f}k$. From
the BCS side to the BEC side, $c_{f}$ evolves from $v_{F}/\sqrt{3}$ to $%
\sqrt{\pi \hbar ^{2}a_{m}n_{m}/m_{f}^{2}}$ \cite{RMP}\cite{Thomas}, where $%
a_{m}=0.6a_{f}$ \cite{Petrov}, and $n_{m}=n_{\uparrow }=n_{\downarrow
}=n_{f} $.

\begin{figure}[t]
\includegraphics[width=3.45 in]{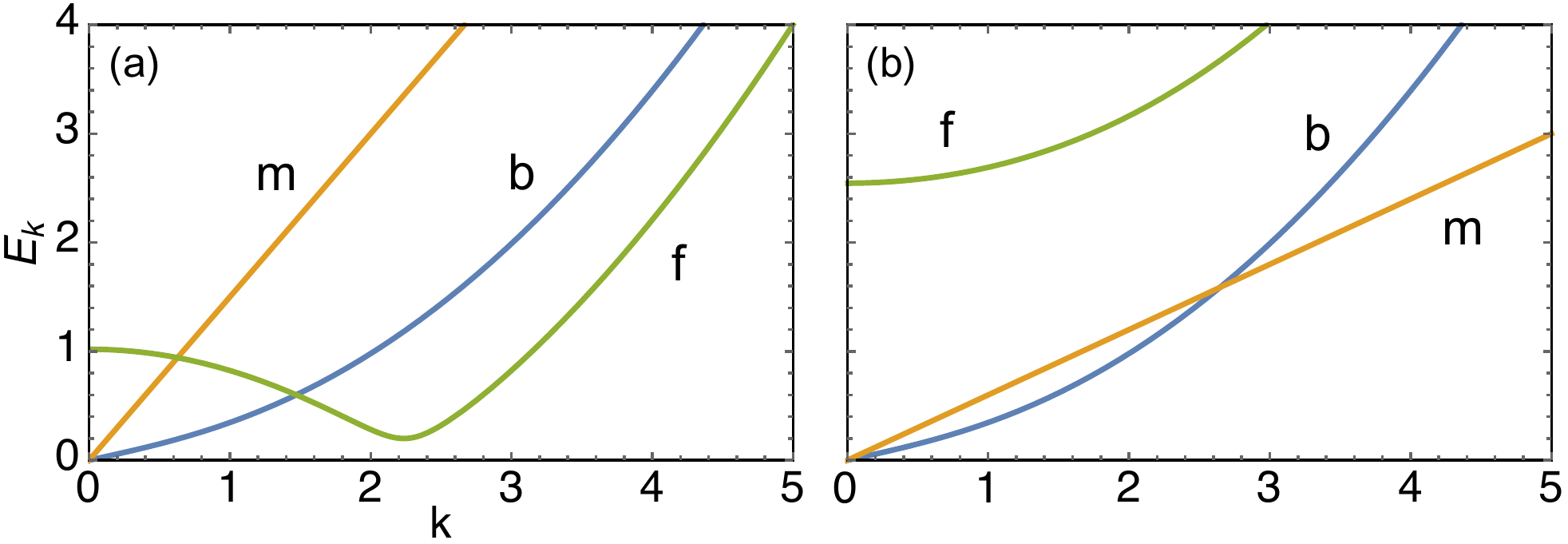}
\caption{Schematic of dispersions of the bosonic Goldstone mode in Bose
superfluid ($E_{\mathbf{k}}^b $), the fermionic pair-breaking mode in Fermi
superfluid($E_{\mathbf{k}}^f $) and the bosonic Goldstone mode in Fermi
superfluid ($E_{\mathbf{k}}^m $). (a) is in the BCS side and (b) is in the
BEC side. }
\label{dis}
\end{figure}

Therefore there are three different excitations in Bose-Fermi superfluid
mixture: the bosonic Goldstone mode in Bose superfluid, whose dispersion is
given by $E_{\mathbf{k}}^{b}$, the fermionic pair-breaking mode in Fermi
superfluid, whose dispersion is given by $E_{\mathbf{k}}^{f}$ and the
bosonic Goldstone mode in Fermi superfluid, whose dispersion is given by $E_{%
\mathbf{k}}^{m}$. At the BCS side, the fermions form Cooper pairs and become
superfluid of BCS type. At the BEC side, the fermions form strongly bound
molecules and become superfluid of BEC type. The dispersions of the three
excitations at different sides are shown respectively in Fig. \ref{dis}(a)
and (b). In the experiment setup, the Bose gas is so dilute that we can take
the free-particle limit, i.e. $E_{\mathbf{k}}^{b}\approx \varepsilon _{%
\mathbf{k}}^{b}$ for the Bogoliubov mode in the Bose superfluid. In this
case, Landau-Beliaev damping of the bosonic mode in Bose superfluid due to
its interaction with itself can be ignored.

\section{Damping at the BCS Side}

Consider the interaction between bosons and fermions. Due to the existence
of boson condensate, $\hat{b}_{0}$ and $\hat{b}_{0}^{\dagger }$ can be
treated as c-numbers, i.e. $\hat{b}_{0}=\hat{b}_{0}^{\dagger }=\sqrt{N_{b}}$%
. Expand the Hamiltonian by the order of $\sqrt{N_{b}}$, we have $\hat{H}%
_{bf}=\hat{H}_{bf}^{(1)}+\hat{H}_{bf}^{(2)}+\hat{H}_{bf}^{(3)}$, where
\begin{align}
\hat{H}_{bf}^{(1)}& =g_{bf}n_{b}\sum_{\mathbf{k},\sigma }\hat{c}_{\mathbf{k}%
,\sigma }^{\dagger }\hat{c}_{\mathbf{k},\sigma }, \\
\hat{H}_{bf}^{(2)}& =\frac{g_{bf}\sqrt{N_{b}}}{V}\sum_{\mathbf{k}\neq 0,%
\mathbf{q},\sigma }(\hat{c}_{\mathbf{q}+\mathbf{k},\sigma }^{\dagger }\hat{c}%
_{\mathbf{q},\sigma }\hat{b}_{\mathbf{k}}+\mathrm{h.c.}), \\
\hat{H}_{bf}^{(3)}& =\frac{g_{bf}}{V}\sum_{\mathbf{k},\mathbf{q}\neq 0,%
\mathbf{p},\sigma }\hat{c}_{\mathbf{p}-\mathbf{q},\sigma }^{\dagger }\hat{c}%
_{\mathbf{p}-\mathbf{k},\sigma }\hat{b}_{\mathbf{q}}^{\dagger }\hat{b}_{%
\mathbf{k}}.
\end{align}%
The leading term $\hat{H}_{bf}^{(1)}$ will only shift the chemical potential
of fermions. The subleading term $\hat{H}_{bf}^{(2)}$ will induce the
damping of boson quasiparticles. The last term $\hat{H}_{bf}^{(3)}$ is a two
particle scattering process, which is less important compared to $\hat{H}%
_{bf}^{(2)}$, so that it can be ignored. We get several damping channels by
expressing $\hat{H}_{bf}^{(2)}$ with mean-field quasiparticle operators: $%
\hat{H}_{bf}^{(2)}\approx \hat{H}_{1}+\hat{H}_{2}+\hat{H}_{3}$. Here
\begin{widetext}
\begin{align}
\hat{H}_1&=\frac{g_{bf}\sqrt{N_{b}}}{V}\sum_{\vect{k},\vect{q},\sigma}(u^{b}_{\vect{k}}-v^{b}_{\vect{k}})(u^{f}_{\vect{q}+\vect{k}}u^{f}_{\vect{q}}-v^{f}_{\vect{q}+\vect{k}}v^{f}_{\vect{q}})\hat{\beta}^\dagger_{\vect{q}+\vect{k}}\hat{\beta}_{\vect{q}}\hat{\alpha}_{\vect{k}}+\mathrm{h.c.} \\
\hat{H}_2&=\frac{g_{bf}\sqrt{N_{b}}}{V}\sum_{\vect{k},\vect{q},\sigma}(u^{b}_{\vect{k}}-v^{b}_{\vect{k}})(u^{f}_{\vect{q}+\vect{k}}u^{f}_{\vect{q}}-v^{f}_{\vect{q}+\vect{k}}v^{f}_{\vect{q}})\hat{\gamma}^\dagger_{\vect{q}+\vect{k}}\hat{\gamma}_{\vect{q}}\hat{\alpha}_{\vect{k}}+\mathrm{h.c.} \\
\hat{H}_3&=\frac{g_{bf}\sqrt{N_{b}}}{V}\sum_{\vect{k},\vect{q},\sigma}(u^{b}_{\vect{k}}-v^{b}_{\vect{k}})(u^f_{\vect{k}-\vect{q}}v^{f}_{\vect{q}}-v^f_{\vect{k}-\vect{q}}u^{f}_{\vect{q}})\hat{\gamma}^\dagger_{\vect{q}}\hat{\beta}^\dagger_{\vect{k}-\vect{q}}\hat{\alpha}_{\vect{k}}+\mathrm{h.c.}
\end{align}
\end{widetext}We have ignored terms like $\hat{\beta}_{\mathbf{q}}\hat{\beta}%
_{\mathbf{q}-\mathbf{k}}\hat{\alpha}_{\mathbf{k}}$ that violate the
conservation of energy, which will not contribute to damping process.

We note that $\hat{H}_{3}$ is the decay process of bosonic mode in Bose
superfluid, which is Beliaev damping. $\hat{H}_{1}$ and $\hat{H}_{2}$ are
the scattering of the bosonic mode in Bose superfluid by thermal
excitations, corresponding to Landau damping. Unlike Landau-Beliaev damping
in single component Bose gas, there exists nonzero damping threshold for
some of the damping channels. While due to the energy gap $\Delta $ in $\hat{%
\beta}_{\mathbf{q}+\mathbf{k}}$ and $\hat{\gamma}_{\mathbf{q}+\mathbf{k}}$.
The damping threshold for $\hat{H}_{3}$ is $\Omega _{3}(\mathbf{k})=\min [E_{%
\mathbf{q}}^{f}+E_{\mathbf{k}-\mathbf{q}}^{f}]>2\Delta $ which is nonzero.
That is to say, for low energy excitations with $E_{\mathbf{k}}^{b}<\Omega
_{3}(\mathbf{k})$, there will be no damping contributed from channel $\hat{H}%
_{3}$. So the Beliaev damping has a critical momentum. The damping threshold
of $\hat{H}_{1}$ and $\hat{H}_{2}$ are the same, which is $\Omega _{1}(%
\mathbf{k})=\Omega _{2}(\mathbf{k})=\min [E_{\mathbf{q}+\mathbf{k}}^{f}-E_{%
\mathbf{q}}^{f}]=0$. So the Landau damping we considered here has no damping
threshold, i.e. the critical momentum for the damping is zero.

As shown in Fig. \ref{dis}(a), there is also a phonon-like bosonic mode of
center-of-mass motion of Cooper pairs in Fermi superfluid. We will discuss
its contribution to the Landau damping later.

\begin{figure}[t]
\includegraphics[width=3.4 in]
{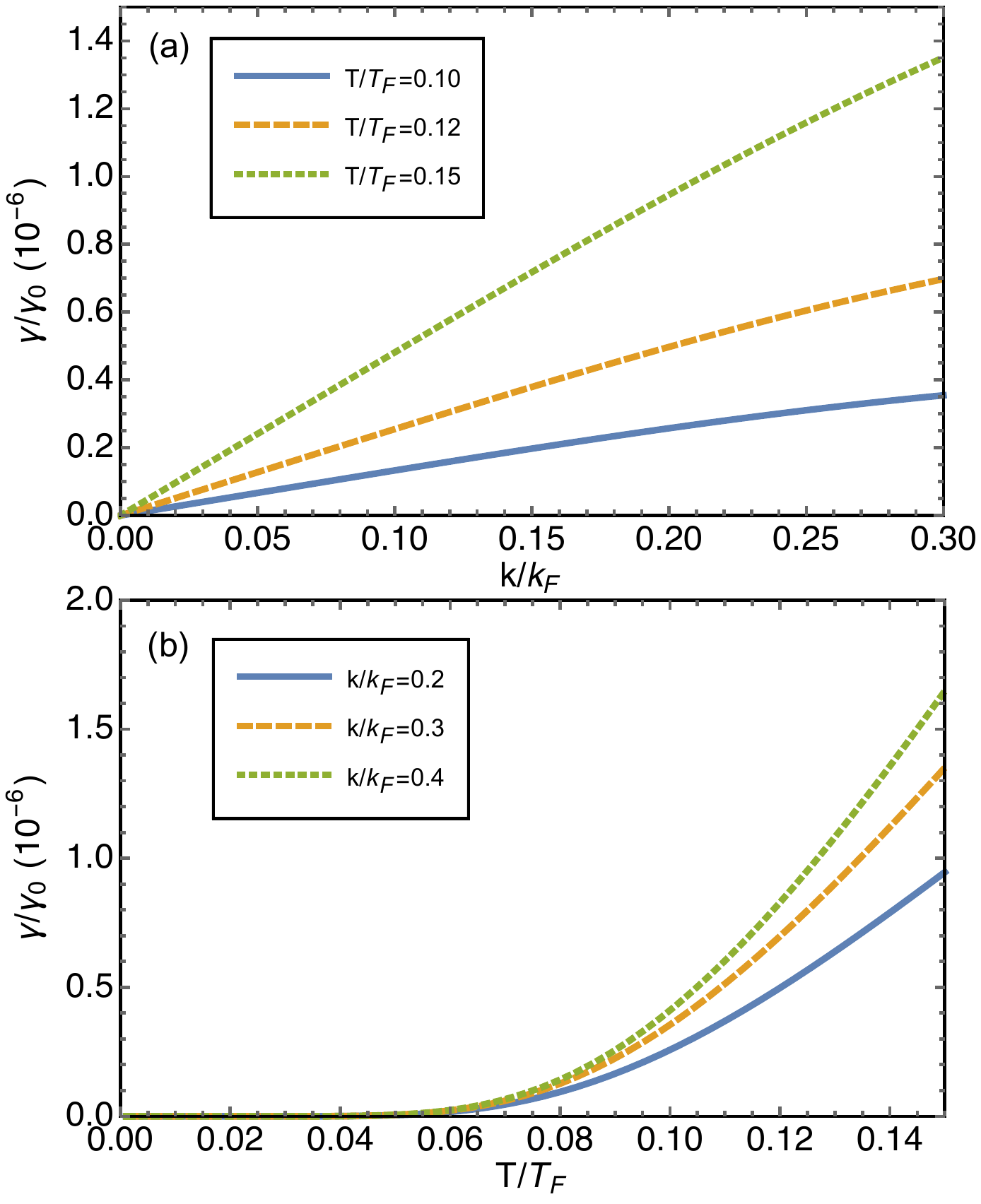}
\caption{Damping rates $\protect\gamma $ in units of $\protect\gamma_0 $ ($%
\protect\gamma_0\equiv E_F/\hbar $) as a function of $k/k_F $ with $%
1/(k_Fa_f)=-0.5 $ in the BCS side. (a) Damping rates as a function of
momentum; (b) Damping rates as a function of temperature. }
\label{BCS}
\end{figure}

Since the interaction between bosons and fermions is weak compared to the
excitation energy in both bosons and fermions, so that we can treat it
perturbatively. According to Fermi's Golden Rule, the rate for the Landau
damping described by $\hat{H}_{1}$ and $\hat{H}_{2}$ is given by
\begin{equation}
\gamma (\mathbf{k})=\frac{2\pi }{\hbar }\sum_{\mathbf{q}}|M_{\mathbf{q}%
\mathbf{k}}|^{2}\delta (E_{\mathbf{k}}^{b}+E_{\mathbf{q}}^{f}-E_{\mathbf{q}+%
\mathbf{k}}^{f})\left[ f(E_{\mathbf{q}}^{f})-f(E_{\mathbf{q}+\mathbf{k}}^{f})%
\right] ,  \label{bcsgdr}
\end{equation}%
where $f(E_{\mathbf{q}}^{f})=[\exp {(E_{\mathbf{q}}^{f}/T)}+1]^{-1}$ is the
Fermi-Dirac distribution function. The matrix element is
\begin{equation}
M_{\mathbf{q}\mathbf{k}}=\frac{2g_{bf}\sqrt{N_{b}}}{V}(u_{\mathbf{k}}^{b}-v_{%
\mathbf{k}}^{b})(u_{\mathbf{q}+\mathbf{k}}^{f}u_{\mathbf{q}}^{f}-v_{\mathbf{q%
}+\mathbf{k}}^{f}v_{\mathbf{q}}^{f}).
\end{equation}%
The damping rate given by \eqref{bcsgdr} is numerically calculated and the
result is plotted in Fig. \ref{BCS}.

In Fig. \ref{BCS}(a), the damping rate is zero for $k=0 $. It is linear in $%
k $ when $k $ is small. The linear behavior is the result of matrix element
and density-of-state of quasiparticles together. The slope is larger at
higher temperature due to more thermal excitations.

What is more interesting is the temperature dependence of Landau damping
rate. As shown in Fig. \ref{BCS}(b), when $T/T_{F}\ll 1$, the damping rate
shows a $e^{-T_{F}/T}$ behavior. This exponential decay is very different
from the power law behavior of Landau damping in dilute Bose gas. Similar as
the damping threshold discussed before, the conservation of energy must be
satisfied in the damping process. Due to the existence of the gap $\Delta $
in $\hat{\beta}_{\mathbf{k}}$ and $\hat{\gamma}_{\mathbf{k}}$, in order for
quasiparticle to be damped, it must absorb a thermal excitation with $E_{%
\mathbf{q}}^{f}\gtrsim \Delta $. However, as temperature tends to zero, the
distribution function $f(E_{\mathbf{q}}^{f})$ tends to be $e^{-\Delta
/(k_{B}T)}$. The number of thermal excitations is exponentially suppressed
at low temperature $k_{B}T<\Delta $. Therefore the damping rate is also
exponentially suppressed. The size of the suppressed region is proportional
to gap. The region will be larger if the BCS gap is tuned to be larger. For
usual Landau damping in dilute Bose gas, thermal excitations are gapless. So
even at low temperature, a number of thermal excitations can be excited and
contribute to damping, leading to a $T^{4}$ power law temperature dependence.

\section{Damping at the BEC Side}

According to the result at the BCS side, the damping rate contributed by the
pair-breaking channel will have a exponentially suppressed region. In the
BEC side the molecule is tightly bounded and the pair-breaking energy $%
\Delta $ is quite high. So this suppressed region is large. This damping
channel can be neglected.

Now the damping channel of center-of-mass motion of Cooper pairs in Fermi
superfluid is important. A comprehensive description of Goldstone mode in
Fermi superfluid and it coupling to Bose superfluid can be obtained from
fluctuation theory of Fermi superfluid \cite{bmode}. Here to expose physics
in a simple way, we treat Fermi superfluid at the BEC side as a molecular
BEC. So the Hamiltonian is given by three parts: $\hat{H}=\hat{H}_{b}+\hat{H}%
_{m}+\hat{H}_{bm}$. $H_{b}$ is the same as before.
\begin{align}
\hat{H}_{m}=& \int d^{3}\mathbf{r}\left\{ \hat{d}^{\dagger }(\mathbf{r})\hat{%
H}_{0,m}\hat{d}(\mathbf{r})+\frac{g_{m}}{2}\hat{d}^{\dagger }(\mathbf{r})%
\hat{d}^{\dagger }(\mathbf{r})\hat{d}(\mathbf{r})\hat{d}(\mathbf{r})\right\}
,  \notag \\
\hat{H}_{bm}=& g_{bm}\int d^{3}\mathbf{r}\hat{d}^{\dagger }(\mathbf{r})\hat{d%
}(\mathbf{r})\hat{b}^{\dagger }(\mathbf{r})\hat{b}(\mathbf{r}),
\end{align}%
where $E_{0,m}=-\hbar ^{2}\nabla ^{2}/(2m_{m})-\mu _{m}$. $%
1/g_m=m_m/(4\pi\hbar^2a_m) $. $m_m=2m_f $ and $a_m=0.6a_f $. For $\hat{H}%
_{m} $, we take the Bogoliubov mean-field theory. The result is
\begin{equation}
\hat{H}_{m}^{\mathrm{MF}}=\sum_{\mathbf{k}}E_{\mathbf{k}}^{m}\hat{\chi} _{%
\mathbf{k}}^{\dagger }\hat{\chi} _{\mathbf{k}},
\end{equation}%
where $E_{\mathbf{k}}^{m}=\sqrt{\varepsilon _{\mathbf{k}}^{m}(\varepsilon _{%
\mathbf{k}}^{m}+2g_{m}n_{m})}$, $\hat{\chi} _{k}$ is the quasiparticle
operator for bosonic Goldstone mode in molecular BEC. $\varepsilon _{\mathbf{%
k}}^{m}=\hbar ^{2}k^{2}/(2m_{m})$ is the kinetic energy of molecules. The
corresponding Bogoliubov transformation of molecule operators is given by $%
d_{\mathbf{k}}=u_{\mathbf{k}}^{b}\hat{\chi} _{\mathbf{k}}-v_{\mathbf{k}}^{b}%
\hat{\chi} _{-\mathbf{k}}^{\dagger }$, where momentum-dependent coefficients
$u_{\mathbf{k}}^{m}$ and $v_{\mathbf{k}}^{m}$ are given by
\begin{equation}
u_{\mathbf{k}}^{m}(v_{\mathbf{k}}^{m})=\sqrt{\frac{1}{2}\left( \frac{%
\varepsilon _{\mathbf{k}}^{m}+g_{m}n_{m}}{E_{\mathbf{k}}^{m}}\pm 1\right) }.
\end{equation}

Due to the existence of condensates, $\hat{b}_{0}$, $\hat{b}_{0}^{\dagger }$%
, $\hat{d}_{0}$ and $\hat{d}_{0}^{\dagger }$ can be replaced by $\hat{b}_{0}=%
\hat{b}_{0}^{\dagger }=\sqrt{N_{b}}$, $\hat{d}_{0}=\hat{d}_{0}^{\dagger }=%
\sqrt{N_{m}}$. To the order of $\sqrt{N_{b}}$ and $\sqrt{N_{m}}$, $\hat{H}%
_{bm}=\hat{H}_{bm}^{(1)}+\hat{H}_{bm}^{(2)}+\hat{H}_{bm}^{(3)}$, where
\begin{align}
\hat{H}_{bm}^{(1)}=& g_{bm}\sum_{\mathbf{k}\neq 0}(n_{m}\hat{b}_{\mathbf{k}%
}^{\dagger }\hat{b}_{\mathbf{k}}+n_{b}\hat{d}_{\mathbf{k}}^{\dagger }\hat{d}%
_{\mathbf{k}}), \\
\hat{H}_{bm}^{(2)}=& \frac{g_{bm}\sqrt{N_{m}}}{V}\sum_{\mathbf{k},\mathbf{q}%
\neq 0}(\hat{d}_{\mathbf{q}}+\hat{d}_{-\mathbf{q}}^{\dagger })\hat{b}_{%
\mathbf{k}+\mathbf{q}}^{\dagger }\hat{b}_{\mathbf{k}}  \notag \\
& +\frac{g_{bm}\sqrt{N_{b}}}{V}\sum_{\mathbf{k},\mathbf{q}\neq 0}(\hat{d}_{%
\mathbf{q}+\mathbf{k}}^{\dagger }\hat{d}_{\mathbf{q}}\hat{b}_{\mathbf{k}}+%
\hat{d}_{\mathbf{q}-\mathbf{k}}^{\dagger }\hat{d}_{\mathbf{q}}\hat{b}_{%
\mathbf{k}}^{\dagger }), \\
\hat{H}_{bm}^{(3)}=& \frac{g_{bm}}{V}\sum_{\mathbf{p},\mathbf{q}\neq 0,%
\mathbf{k}\neq \mathbf{q}}\hat{d}_{\mathbf{q}-\mathbf{k}}^{\dagger }\hat{d}_{%
\mathbf{q}}\hat{b}_{\mathbf{p}+\mathbf{k}}^{\dagger }\hat{b}_{\mathbf{p}}.
\end{align}%
The leading term $\hat{H}_{bm}^{(1)}$ will only modify the chemical
potential. The subleading term $\hat{H}_{bm}^{(2)}$ will contribute to the
damping. The last term $\hat{H}_{bf}^{(3)}$ is a two particle scattering
process, which is less important compared to the $\hat{H}_{bf}^{(2)}$ and
can be ignored. Express $\hat{H}_{bm}^{(2)}$ with mean-field quasiparticle
operators, we get damping channels $\hat{H}_{bm}^{(2)}\approx \hat{H}_{1}+%
\hat{H}_{2}+\hat{H}_{3}+\hat{H}_{4}$, where
\begin{widetext}
\begin{align}
\hat{H}_{1}&=\frac{g_{bm}\sqrt{N_{m}}}{V}\sum_{\vect{k},\vect{q}}(u_{\vect{q}}^{m}-v_{\vect{q}}^{m})(u_{\vect{k}-\vect{q}}^{b}u_{\vect{k}}^{b}+v_{\vect{k}-\vect{q}}^{b}v_{\vect{k}}^{b})\hat{\chi}^\dagger_{\vect{q}}\hat{\alpha}^\dagger_{\vect{k}-\vect{q}}\hat{\alpha}_{\vect{k}}+\mathrm{h.c.},\\
\hat{H}_{2}&=-\frac{g_{bm}\sqrt{N_{b}}}{V}\sum_{\vect{k},\vect{q}}(u_{\vect{k}}^{b}-v_{\vect{k}}^{b})u_{\vect{k}-\vect{q}}^mv_{\vect{q}}^{m}\hat{\chi}^\dagger_{\vect{q}}\hat{\chi}^\dagger_{\vect{k}-\vect{q}}\hat{\alpha}_{\vect{k}}+\mathrm{h.c.}, \\
\hat{H}_{3}&=-\frac{g_{bm}\sqrt{N_{m}}}{V}\sum_{\vect{k},\vect{q}}(u_{\vect{q}+\vect{k}}^{m}-v_{\vect{q}+\vect{k}}^{m})u_{\vect{k}}^{b}v_{\vect{q}}^{b}\hat{\chi}^\dagger_{\vect{q}+\vect{k}}\hat{\alpha}_{\vect{q}}\hat{\alpha}_{\vect{k}}+\mathrm{h.c.}, \\
\hat{H}_{4}&=\frac{g_{bm}\sqrt{N_{b}}}{V}\sum_{\vect{k},\vect{q}}(u_{\vect{k}}^{b}-v_{\vect{k}}^{b})(u_{\vect{q}+\vect{k}}^{m}u_{\vect{q}}^{m}+v_{\vect{q}+\vect{k}}^{m}v_{\vect{q}}^{m})\hat{\chi}^\dagger_{\vect{q}+\vect{k}}\hat{\chi}_{\vect{q}}\hat{\alpha}_{\vect{k}}+\mathrm{h.c.},
\end{align}
\end{widetext}We have ignored terms like $\hat{\chi}_{-\mathbf{q}}\hat{\alpha%
}_{\mathbf{q}+\mathbf{k}}\hat{\alpha}_{\mathbf{k}}$ that violates the
conservation of energy that will not contribute to damping process.

Here $\hat{H}_{2}$ is the decay of bosonic mode in Bose superfluid, which is
Beliaev damping. $\hat{H}_{3}$ and $\hat{H}_{4}$ are scattering by thermal
excitations, which are Landau damping. Special attention should be paid on $%
\hat{H}_{1}$: $\hat{\chi}_{\mathbf{q}}^{\dagger }\hat{\alpha}_{\mathbf{k}-%
\mathbf{q}}^{\dagger }\hat{\alpha}_{\mathbf{k}}$ gives Beliaev damping,
while its Hermitian conjugation $\hat{\alpha}_{\mathbf{q}+\mathbf{k}%
}^{\dagger }\hat{\chi}_{\mathbf{q}}\hat{\alpha}_{\mathbf{k}}$ can contribute
to Landau damping. Similar as the BCS side, $\hat{\alpha}_{\mathbf{k}}$ and $%
\hat{\chi}_{\mathbf{k}}$ are different quasiparticles with different
dispersions. To satisfy conservation of energy and momentum, the Beliaev
damping channels $\hat{H}_{1}$ and $\hat{H}_{2}$ have damping thresholds for
low-energy excitations:
\begin{align}
\Omega _{1}(\mathbf{k})& =%
\begin{cases}
k^{2}/(2m_{b}) & k<m_{b}c_{m}, \\
c_{m}k-m_{b}c_{m}^{2}/2 & k>m_{b}c_{m},%
\end{cases}
\notag \\
\Omega _{2}(\mathbf{k})& =c_{m}k,
\end{align}%
respectively. The Landau damping channels have no damping threshold, i.e.
have a zero critical momentum.

The matrix elements of Landau damping are given by:
\begin{align}
M_{1,\mathbf{q}\mathbf{k}}& =\frac{g_{bm}\sqrt{N_{m}}}{V}(u_{\mathbf{q}%
}^{m}-v_{\mathbf{q}}^{m})(u_{\mathbf{k}-\mathbf{q}}^{b}u_{\mathbf{k}}^{b}+v_{%
\mathbf{k}-\mathbf{q}}^{b}v_{\mathbf{k}}^{b}),  \notag \\
M_{3,\mathbf{q}\mathbf{k}}& =-\frac{g_{bm}\sqrt{N_{m}}}{V}(u_{\mathbf{q}+%
\mathbf{k}}^{m}-v_{\mathbf{q}+\mathbf{k}}^{m})(u_{\mathbf{k}}^{b}v_{\mathbf{q%
}}^{b}+v_{\mathbf{k}}^{b}u_{\mathbf{q}}^{b}),  \notag \\
M_{4,\mathbf{q}\mathbf{k}}& =\frac{g_{bm}\sqrt{N_{b}}}{V}(u_{\mathbf{k}%
}^{b}-v_{\mathbf{k}}^{b})(u_{\mathbf{q}+\mathbf{k}}^{m}u_{\mathbf{q}}^{m}+v_{%
\mathbf{q}+\mathbf{k}}^{m}v_{\mathbf{q}}^{m}).
\end{align}%
As mentioned before, we only consider the free-particle-limit for Goldstone
mode in Bose superfluid. So $u_{\mathbf{k}}^{b}$ and $v_{\mathbf{k}}^{b}$
can be simplified as $u_{\mathbf{k}}^{b}\approx 1$, $v_{\mathbf{k}%
}^{b}\approx 0$. We immediately get that $M_{1,\mathbf{q}\mathbf{k}}=M_{3,%
\mathbf{q}\mathbf{k}}=0$. Physically, this is because in the free-boson
limit $\hat{\alpha}_{\mathbf{k}}\approx \hat{b}_{\mathbf{k}}$. Thus $\hat{H}%
_{1}$ and $\hat{H}_{3}$ violate the conservation of number of bosons.
Therefore we only consider damping channel $\hat{H}_{4}$.

According to Fermi's Golden Rule, this Landau damping rate is given by
\begin{equation}
\gamma (\mathbf{k})=\frac{2\pi }{\hbar }\sum_{\mathbf{q}}|M_{\mathbf{q}%
\mathbf{k}}|^{2}\delta (E_{\mathbf{k}}^{b}+E_{\mathbf{q}-\mathbf{k}}^{m}-E_{%
\mathbf{q}}^{m})\left[ f(E_{\mathbf{q}-\mathbf{k}}^{m})-f(E_{\mathbf{q}}^{m})%
\right] ,  \label{becgdr}
\end{equation}%
where $f(E_{\mathbf{q}}^{m})=[\exp {(E_{\mathbf{q}}^{m}/T)}-1]^{-1}$ is the
Bose-Einstein distribution function. The conservation of energy $E_{\mathbf{k%
}}^{b}+E_{\mathbf{q}-\mathbf{k}}^{m}-E_{\mathbf{q}}^{m}=0 $ will give a
momentum lower bound $q_{c}(\mathbf{k})$ for the thermal excitations, which
is
\begin{equation}
q_{c}(\mathbf{k})=\frac{k}{2}\left( 1-\frac{\hbar k}{2m_{b}c_{m}}\right).
\label{lowerbound}
\end{equation}%
Only thermal excitations with $q>q_{c}(\mathbf{k})$ will contribute to the
damping. The physical reason for this lower bound can be understood similar
as the damping threshold as following: for infinitesimal $\mathbf{q}$, the
conservation of energy and momentum cannot be satisfied simultaneously,
because $\hat{\chi} _{\mathbf{k}}$ are quasiparticles with different
dispersions from $\hat{\alpha} _{\mathbf{k}}$. To be more specific, $\hat{%
\chi}_{\mathbf{k}}$ has a large velocity and has higher energy compared to $%
\hat{\alpha} _{\mathbf{k}}$.

\begin{figure}[t]
\includegraphics[width=3.4 in]{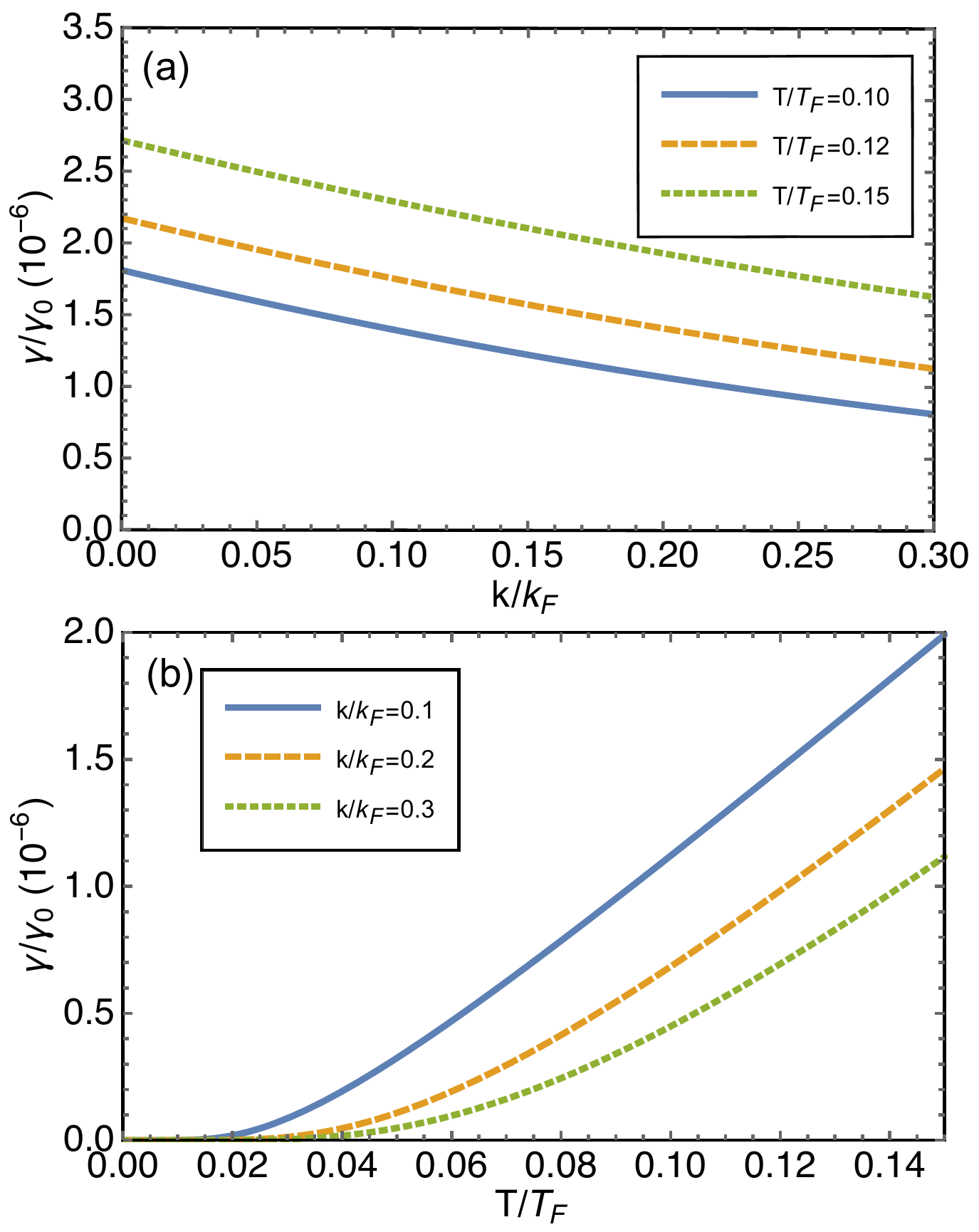}
\caption{Damping rates $\protect\gamma $ in units of $\protect\gamma_0 $ ($%
\protect\gamma_0\equiv E_F/\hbar $) as a function of $k/k_F $ with $%
1/(k_Fa_f)=0.5 $ for the BEC side. (a) Damping rates as a function of
momentum; (b) Damping rates as a function of temperature. }
\label{BEC}
\end{figure}

The damping rate given by \eqref{becgdr} is numerically calculated and the
result is plotted in Fig. \ref{BEC}. In Fig. \ref{BEC}(a), the damping rate
is nonzero at $k=0$. At leading order, the damping rate decreases with
momentum. This can be understood as the competition of matrix element and
the density-of-state of the quasiparticles. The leading and the subleading
term of $\gamma (\mathbf{k})$ can be calculated analytically by taking the
asymptotic expansion of matrix element. The result is
\begin{equation}
\gamma (\mathbf{k})=\frac{g_{bm}^{2}n_{b}m_{m}^{2}}{4\pi \hbar ^{3}m_{b}}%
\left[ \frac{2m_{m}c_{m}T}{\hbar T_{F}}-\left( \frac{1}{2}+\frac{m_{m}T}{%
m_{b}T_{F}}\right) k\right] +\mathcal{O}(k^{2}).
\end{equation}%
The intercept at $k=0$ is proportional to $T/T_{F}$. This is reasonable
because when $T=0$, there are no thermal excitations. The subleading term is
always negative, which agrees with the numerical calculations, see Fig. \ref%
{BEC}(a).

The temperature dependence of the damping rate is plottd in Fig. \ref{BEC}%
(b). We note that $\gamma (T/T_{F})$ is linear in $T/T_{F}$ at high
temperature. At low temperature, when $T/T_{F}\ll 1$, $\gamma
(T/T_{F})\propto e^{-T_{F}/T}$. So at the BEC side, the temperature
dependence also shows an exponentially decay behavior, which is similar to
the BCS side. However, the physical origin of this exponentially suppression
is not the gap, but the momentum lower bound. Since only excitations with $%
q>q_{c}$ can contribute to the damping, there is an energy lower bound $%
E_{c}=\hbar c_{m}q_{c}$ for thermal excitations to involve into Landau
damping process. For low temperature $k_{B}T<E_{c}$, the involved thermal
excitations are exponentially suppressed, leading to an exponentially
suppressed damping rate. The size of this suppressed region is proportional
to energy lower bound $E_{c}$. According to \eqref{lowerbound}, larger the $%
k $, larger the $E_{c}$, hence larger the suppressed region. This is agreed
with Fig. \ref{BEC}(b) and is also a distinct feature from the BCS side. For
the Landau damping in single component BEC, the quasiparticles are also
gapless. But there is only one kind of quasiparticle and thus no energy
lower bound. The conservation of energy and momentum can always be satisfied
there. As a result, there is no exponential suppressed region.

Now we would like to point out that the Landau damping channels via
interacting with the Goldstone mode in Fermi superfluid also exist at the
BCS side. Although the detail of these channels cannot be covered at
mean-field level, they must obey the conservation of energy and momentum.
The velocity of Goldstone mode in Fermi superfluid at the BCS side is about $%
v_{F}/\sqrt{3}$ and is quite large. So based on our previous analysis, the
corresponding energy lower bound $\hbar c_{m}q_{c}$ is also very large at
the BCS side, such that the damping process via those channels are highly
suppressed in a large temperature range. It can be ignored comparing to the
damping channels via interacting with the fermionic pair breaking modes at
the BCS side.

At the end of this section, we should emphasize that our analysis of
temperature dependence in Landau damping is only viable at a relative low
temperature, since both the gap of the Fermi superfluid and the condensate
fraction in Bose superfluid will decrease with the increasing of
temperature. A comprehensive self-consistent analysis of temperature
dependence should take the changing of the gap and the condensate fraction
into account. However, at low temperature both the gap and the condensate
fraction are not sensitive to the temperature, so our result is still
reasonable in this region.

\section{Summary}

In summary, we have studied Landau damping in Bose-Fermi superfluid mixture
at finite temperature. Unlike Beliaev damping in Bose-Fermi superfluid
mixture at zero temperature, Landau damping has no critical momentum, since
any quasiparticle with infinitesimal momentum can be damped by absorbing a
thermal quasiparticle. However, due to energy-momentum conservation, thermal
excitations in Fermi superfluid will be involved into Landau process only if
their energy is larger than a lower bound $E_{c}$. When the temperature is
lower than $E_{c}$, Landau damping rate will be exponentially suppressed.
For fermionic quasiparticles in Fermi superfluid, the lower bound is
determined by pair-breaking gap $\Delta $. For bosonic excitations, the
lower bound is determined by $\hbar c_{m}q_{c}$. This exponential
suppression is quite different from the power law temperature dependence of
Landau damping in single component BEC. The reason behind is that in
Bose-Fermi superfluid mixture, quasiparticle in Bose superfluid is damped by
coupling to thermal excitations in Fermi superfluids with totally different
dispersion. In single component BEC, quasiparticle is damped by interacting
with itself, so that energy-momentum conservation can always be satisfied.
Therefore there will be no energy lower bound and no exponential
suppression. In dipolar Bose gas, there is also an exponential suppression
region for Landau damping \cite{dampdipole}. That is due to its unusual
quasiparticle dispersion with maxon-roton structure.

In principle, both damping channels contributed from bosonic quasiparticles
and fermionic quasiparticles in Fermi superfluid exist at both sides.
However, at the BCS side, we have $\Delta \ll \hbar c_{m}q_{c}$, so that
Landau damping is dominated by fermionic excitations in Fermi superfluid. At
the BEC side, the situation is opposite, $\Delta \gg \hbar c_{m}q_{c}$. So
bosonic quasiparticles in Fermi superfluid dominate Landau process. This
result is similar to the case of zero-temperature Beliaev damping. The
different dominated low energy quasiparticles also lead to totally distinct
momentum dependence of damping rate at the BCS side and the BEC side.

In the ENS experiment \cite{ENS}, damping of dipole modes has a critical
momentum even the experiment is done at finite temperature. Based on our
previous analysis, this phenomenon can be understood as following: the
temperature of the experiment is sufficient low that Landau damping with
zero critical momentum is highly suppressed. Then Beliaev damping with
nonzero critical momentum will show up \cite{ZhengWei}. Our prediction of
momentum dependence of Landau damping could be observed in the same
experiment setup at relative high temperature, when Landau damping will
dominate over Beliaev damping. No critical momentum will be observed there.

\textit{Acknowledgment}: We wish to thank Hui Zhai for helpful discussions.
This work is supported by Tsinghua University Initiative Scientific Research
Program, NSFC Grant No. 11174176, and NKBRSFC under Grant No. 2011CB921500.


\begin{thebibliography}{99}
\bibitem{Stringaribook} L. P. Pitaveski and S. Stringari, \textit{%
Bose-Einstein Condensation}, (Oxford University Press, New York, 2003),
Chapter 6.

\bibitem{Pethick} C. J. Pethick and H. Smith, \textit{Bose-Einstein
Condensation in Dilute Gases}, (Cambridge University Press, New York, 2002),
Chapter 10.

\bibitem{Landau} L. D. Landau, J. Phys. USSR 10 (1946).

\bibitem{Beliaev} S. T. Beliaev, Soviet Phys. JETP \textbf{34}, 299 (1958);

\bibitem{LandauBaliaev} P. C. Hohenberg, P. C. Martin, Ann. Phys. (NY)
\textbf{34}, 291 (1965); P. Szepfalusy, I. Kondor, Ann. Phys. (NY) \textbf{82%
}, 1 (1974).

\bibitem{Vincent} W. V. Liu, Phys. Rev. Lett. \textbf{79}, 4056 (1997);

\bibitem{Stringari2} L. P. Pitaveski and S. Stringari, Phys. Lett. A \textbf{%
235, }398 (1997); S. Giorgini, Phys. Rev. A \textbf{57}, 2949 (1998).

\bibitem{exp2} N. Katz, J. Steinhauer, R. Ozeri, and N. Davidson, Phys. Rev.
Letts. \textbf{89}, 220401 (2002).

\bibitem{exp1} D. S. Jin, J. R. Ensher, M. R. Matthews, C. E. Wieman, and E.
A. Cornell, Phys. Rev. Lett. \textbf{77}, 420 (1996); D. S. Jin, M. R.
Matthews, J. R. Ensher, C. E. Wieman, and E. A. Cornell, Phys. Rev. Lett.
\textbf{78}, 764 (1997); M.-O. Mewes, M. R. Andrews, N. J. van Druten, D. M.
Kurn, D. S. Durfee, C. G. Townsend, and W. Ketterle, Phys. Rev. Lett.
\textbf{77}, 988 (1996); N. Katz, J. Steinhauer, R. Ozeri, and N. Davidson,
Phys. Rev. Letts. \textbf{89}, 220401 (2002).

\bibitem{Gora} P. O. Fedichev, G. V. Shlyapnikov and J. T. M. Walraven,
Phys. Rev. Lett. \textbf{80}, 2269 (1998).

\bibitem{dampdipole} S. S. Natu, S. Das Sarma, Phys. Rev. A \textbf{88,}
031604 (R) (2013); S. S. Natu, R. M. Wilson, Phys. Rev. A \textbf{88,}
063638 (2013).

\bibitem{dampmix} D. H. Santamore, S. Gaudio, E. Timmermans, Phys. Rev.
Lett. \textbf{93}, 250402 (2004); S. K. Yip, Phys. Rev. A \textbf{64},
023609 (2001); D. H. Santamore, E. Timmermans, Phys. Rev. A \textbf{72},
053601 (2005); X.-J. Liu and H. Hu, Phys. Rev. A \textbf{68}, 033613 (2003);
J. H. Pixley, Xiaopeng Li, and S. Das Sarma, arXiv:1501.05015v1.

\bibitem{ENS} I. Ferrier-Barbut, M. Delehaye, S. Laurent, A. T. Grier, M.
Pierce, B. S. Rem, F. Chevy, C. Salomon, Science, 345, 1035 (2014).

\bibitem{pre} Previously, there are few theoretical studies of Bose-Fermi
superfluid mixture. I. M. Khalatnikov, ZhETF Pis. Red. \textbf{17}, No. 9,
534 (1973); A. F. Andreev and E. P. Bashkin, Zh. Eksp. Teor. Fiz. \textbf{69}%
, 319 (1975); H, Shibata, N, Yokoshi, and S, Kurihara, Phys. Rev. A \textbf{%
75}, 053615 (2007); S. K. Adhikari and L. Salasnich, Phys. Rev. A \textbf{78}%
, 043616 (2008); B. Ramachandhran, S. G. Bhongale, and H. Pu, Phys. Rev. A
\textbf{83}, 033607 (2011);

\bibitem{Stringari} T. Ozawa, A. Recati, S. Stringari, Phys. Rev. A \textbf{%
90}, 043608 (2014).

\bibitem{ad} R. Zhang, W. Zhang, H. Zhai and P. Zhang, Phys. Rev. A \textbf{%
90}, 063614 (2014); X. Cui, Phys. Rev. A \textbf{90}, 041603(R) (2014).

\bibitem{ZhengWei} Wei Zheng and Hui Zhai, Phys. Rev. Lett. \textbf{113},
265304 (2014).

\bibitem{rotMix} Linghua Wen and Jinghong Li, Phys. Rev. A \textbf{90},
053621 (2014).

\bibitem{Bruun} J. J. Kinnunen and G. M. Bruun, arXiv: 1502.00402v1.

\bibitem{FGoldstone} P. W. Anderson, Phys. Rev. \textbf{112}, 1900 (1958).

\bibitem{RMP} S. Giorgini, L. P. Pitaevskii, and S. Stringari, Rev. Mod.
Phys. \textbf{80}, 1215 (2008).

\bibitem{Thomas} J. Joseph, B. Clancy, L. Luo, J. Kinast, A. Turlapov, and
J. E. Thomas, Phys. Rev. Lett. \textbf{98}, 170401, (2007).

\bibitem{Petrov} D. S. Petrov, C. Salomon, and G. V. Shlyapnikov, Phys. Rev.
A \textbf{71}, 012708 (2005).

\bibitem{bmode} C. A. R. S\'{a} de Melo, Mohit Randeria, and Jan R.
Engelbrecht, Phys. Pev. Lett. \textbf{71}, 3202 (1993); Jan R. Engelbrecht,
Mohit Randeria, and C. A. R. S\'{a} de Melo, Phys. Rev. B \textbf{55}, 15153
(1997); Roberto B. Diener, Rajdeep Sensarma, and Mohit Randeria, Phys. Rev.
A \textbf{77}, 023626 (2008).
\end{thebibliography}
\end{document}